\documentclass{emulateapj}
\usepackage{natbib}
\usepackage{amssymb}
\usepackage{amsmath}

\slugcomment{Revised manuscript for publication in \emph{ApJ}, \today}

\shorttitle{Chemical Timescales on Highly Eccentric Exoplanets}

\shortauthors{Visscher}

\begin{document}

\title{Chemical Timescales in the Atmospheres of Highly Eccentric Exoplanets}

\author{Channon Visscher}
\affil{Department of Space Studies, Southwest Research Institute, Boulder, CO, 80302, USA}

\begin{abstract}
Close-in exoplanets with highly eccentric orbits are subject to large variations in incoming stellar flux between periapse and apoapse.  These variations may lead to large swings in atmospheric temperature, which in turn may cause changes in the chemistry of the atmosphere from higher CO abundances at periapse to higher CH$_{4}$ abundances at apoapse.  Here we examine chemical timescales for CO$\rightleftarrows$CH$_{4}$ interconversion compared to orbital timescales and vertical mixing timescales for the highly eccentric exoplanets HAT-P-2b and CoRoT-10b.  As exoplanet atmospheres cool, the chemical timescales for CO$\rightleftarrows$CH$_{4}$ tend to exceed orbital and/or vertical mixing timescales, leading to quenching.  The relative roles of orbit-induced thermal quenching and vertical quenching depend upon mixing timescales relative to orbital timescales.  For both HAT-P-2b and CoRoT-10b, vertical quenching will determine disequilibrium CO$\rightleftarrows$CH$_{4}$ chemistry at faster vertical mixing rates ($K_{zz}>10^7$ cm$^2$ s$^{-1}$), whereas orbit-induced thermal quenching may play a significant role at slower mixing rates ($K_{zz}<10^7$ cm$^2$ s$^{-1}$).  The general abundance and chemical timescale results -- calculated as a function of pressure, temperature, and metallicity -- can be applied for different atmospheric profiles in order to estimate the quench level and disequilibrium abundances of CO and CH$_{4}$ on hydrogen-dominated exoplanets.  Observations of CO and CH$_{4}$ on highly eccentric exoplanets may yield important clues to the chemical and dynamical properties of their atmospheres.
\end{abstract}

\keywords{planetary systems --- 
planets and satellites: atmospheres --- 
planets and satellites: composition --- 
planets and satellites: individual (HAT-P-2b, CoRoT-10b)}

\section{Introduction}

For planets with high eccentricity, the large variations in flux received from their host stars may yield substantial variations in atmospheric temperature and dynamical behavior during the course of an orbit \citep[][]{langton2008,laughlin2009,iro2010,cowan2011,kane2011,rauscher2012,lewis2012}.  In some cases, orbit-induced temperature variations may be large enough to produce a significant shifts in the chemical behavior of the planet.  For example, the swing between high atmospheric temperatures at periapse and lower atmospheric temperatures at apoapse may shift equilibrium chemistry predictions from relatively higher CO abundances at periapse (toward a CO-dominated atmosphere) to relatively higher CH$_{4}$ abundances at apoapse (toward a CH$_{4}$-dominated atmosphere).

Large changes in the abundances of CO and CH$_{4}$ are of particular interest because these compounds strongly influence the spectral properties of exoplanet atmospheres \citep[e.g.,][]{seager2000,burrows2005apjl,charbonneau2007,charbonneau2008,fortney2007,barman2008,swain2008,swain2009apj,swain2009apjl,swain2010,desert2009,madhusudhan2009,madhusudhan2011,madhusudhan2011nature,Stevenson2010,tinetti2010,tinetti2010faraday,beaulieu2011,knutson2011,lee2012,shabram2011,waldmann2012}.  Phase-dependent variations in the planetary spectrum \citep[e.g.,][]{barman2005,fortney2006dyn,knutson2007nature,knutson2012,showman2008,showman2009,cowan2011,lewis2012} may therefore also reflect changes in chemical composition of highly eccentric transiting exoplanets (see Table \ref{tab: exoplanets}), particularly at wavelengths sensitive to CO and CH$_{4}$ \citep[e.g.,][]{lewis2012}

However, the extent of temperature-dependent variations in carbon chemistry throughout the orbit -- and whether equilibrium chemistry at a given altitude prevails over orbital timescales -- also depends upon the rate of CO$\rightleftarrows$CH$_{4}$ interconversion relative to the time elapsed between periapse and apoapse.  
Thermochemical equilibrium can be maintained throughout the orbit only if chemical timescales are less than orbital timescales.  To study this effect, \citet{iro2010} estimated the CO$\rightleftarrows$CH$_{4}$ interconversion timescale for HD 80606b and HD 17156b using the kinetics of \citet{bezard2002} and found that orbital timescales are generally much shorter than chemical timescales -- indicative of disequilibrium chemistry -- at pressure levels where orbit-induced temperature variations are expected to be significant.  This behavior will occur on objects which are expected to have relatively low atmospheric temperatures (and therefore sluggish reaction kinetics) near apoapse, leading to orbit-induced thermal quenching wherein an equilibrium composition achieved near periapse survives to become a disequilibrium composition at apoapse.  

The observed properties of highly eccentric exoplanets are subject to numerous variables including thermochemical and photochemical reaction rates, convective transport, and horizontal dynamical and radiative timescales.  Here, we focus specifically on the role of thermochemical quench chemistry in response to vertical transport and eccentricity-induced atmospheric temperature variations, as this behavior may strongly influence observable CO and CH$_{4}$ abundances (even if photolysis occurs at higher altitudes).  For simplicity, the relevant timescale for the temperature variation is taken to be $0.5p$ (where $p$ is the orbital period), which describes the time elapsed between temperature swings at periapse and apoapse.  We first calculate the abundances of CO and CH$_{4}$ and CO$\rightleftarrows$CH$_{4}$ chemical timescales as a function of pressure, temperature, and metallicity in a solar-composition gas, using updated kinetics for CO$\rightleftarrows$CH$_{4}$ interconversion \citep{visscher2010icarus,moses2011,visscher2011}.  The results are then compared to orbital and mixing timescales and pressure-temperature profiles of individual objects to estimate the quench levels and abundances of CO and the CH$_{4}$ as disequilibrium species.  The primary objective of this study is to discuss a convenient method for exploring thermal quenching processes in the CO$\rightleftarrows$CH$_{4}$ system in exoplanet atmospheres.  Although we focus on CO-abundant HAT-P-2b and CH$_{4}$-abundant CoRoT-10b as specific examples, the abundance and timescale results presented here may in principle be applied to any H$_{2}$-dominated substellar object that is subject to orbit-induced temperature variations.
 
\begin{deluxetable}{lccr@{.}lccc}
\tablecaption{Highly-Eccentric ($e>0.3$) Transiting Exoplanets}
\tablewidth{0pt}
\tablehead{
\colhead{Object} & $a$(AU) & $e$ & \multicolumn{2}{c}{$p$(days)} & $p_{s}$(days) & $M{_\textrm{J}}$ & $R_{\textrm{J}}$}
\startdata
HD 80606b & 0.447 & 0.934 & 111&44 & 1.7 & 3.9 & 1.03\\
HD 17156b & 0.163 & 0.682 & 21&22  & 3.6 & 3.3 & 1.02\\
CoRoT-10b & 0.105 & 0.530 & 13&24  & 4.3 & 2.8 & 0.97\\
HAT-P-2b  & 0.068 & 0.517 & 5&63   & 1.9 & 8.9 & 1.16\\
HAT-P-34b & 0.068 & 0.440 & 5&45   & 2.4 & 3.4 & 1.20\\
HAT-P-17b & 0.088 & 0.346 & 10&34  & 5.9 & 0.5 & 1.01\\
WASP-8b   & 0.080 & 0.310 & 8&16   & 5.1 & 2.1 & 1.04\\[-1.5mm]
\enddata
\tablecomments{Data from \citet{wright2011}. The pseudo-synchronous rotation period $p_{s}$ was calculated using the expressions of \citet{hut1981} as presented in \citet{iro2010}.}
\label{tab: exoplanets}
\end{deluxetable}
\section{Chemical Model}
\subsection{Equilibrium Chemistry and Abundances}

\begin{figure*}
\begin{center}
\scalebox{0.45}{\includegraphics[angle=0]{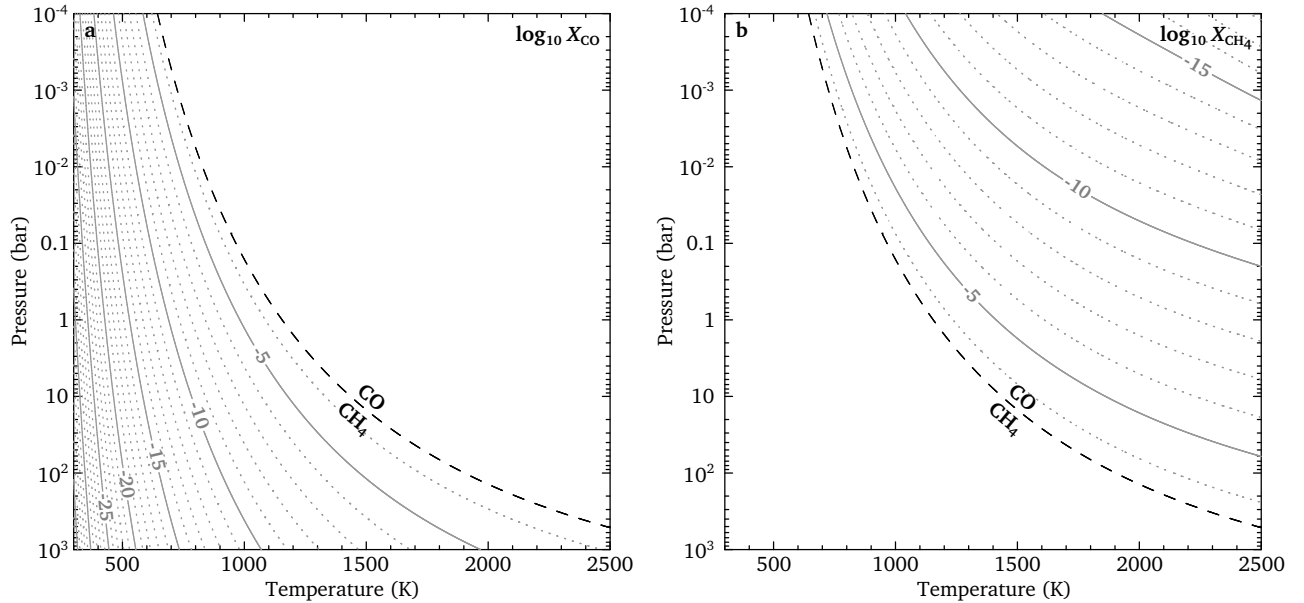}} \end{center} \caption[Carbon Abundances]{Equilibrium mole fraction abundance contours (on a logarithmic scale) for (a) CO and (b) CH$_{4}$ in a solar-composition gas at pressures and temperatures relevant for extrasolar giant planet atmospheres.  In each panel, the dashed line denotes the CO = CH$_{4}$ equal-abundance boundary where $X_{\textrm{CO}} = X_{\textrm{CH}_{4}}\approx 0.5X_{\Sigma\textrm{C}}$.  Figure after \citet{lodders2002}}
\label{fig: abundances} 
\end{figure*}

We develop a gas-phase chemical model based upon the thermochemical equilibrium models of \citet{fegley1994}, \citet{lodders2002}, and \citet{visscher2006,visscher2010apj}.  As in \citet{lodders2002}, we examine H-C-N-O chemistry over a range of pressures, temperatures, and metallicities  so that the model results may be applied to a variety of substellar objects.  Solar-composition elemental abundances are taken from \citet{lodders2010}, and we consider the removal of $\sim21\%$ of oxygen into rock-forming elements \citep{lodders2004apj,visscher2005}.  Figure \ref{fig: abundances} shows the equilibrium mole fraction abundances of CO and CH$_{4}$ as a function of pressure and temperature in a solar-metallicity gas.  Also shown is the CO=CH$_{4}$ equal abundance boundary ($X_{\textrm{CO}}=X_{\textrm{CH}_{4}}=0.5X_{\Sigma\textrm{C}}$),  approximated by
\begin{equation}
\log P \approx 5.05 - 5807.5/T + 0.5[\textrm{Fe/H}],
\end{equation}
for $P$ in bars and $T$ in kelvins.  At pressures and temperatures where methane is the dominant C-bearing gas (higher $P$, lower $T$), the mole fraction abundances of CH$_{4}$ and CO are given by
\begin{align}
X_{\textrm{CH}_{4}} &\approx X^{*}_{\Sigma\textrm{C}}m,\\
X_{\textrm{CO}} &\approx X^{*}_{\textrm{CO}}(P,T)m^{2},
\end{align}
where $X^{*}_{\Sigma\textrm{C}}$ is the total carbon abundance in a solar-metallicity gas \citep[$\sim 4.5\times10^{-4}$;][]{lodders2010}, $X^{*}_{\textrm{CO}}(P,T)$ is the solar-metallicity abundance of CO as a function of pressure and temperature as plotted in Figure \ref{fig: abundances}a, and $m$ is the metallicity factor defined by $\log m = [\textrm{Fe/H}]$ \citep{lodders2002,visscher2006}.

At pressures and temperatures where carbon monoxide is the dominant C-bearing gas (lower $P$, higher $T$), the mole fraction abundances of CO and CH$_{4}$ are given by 
\begin{align}
X_{\textrm{CO}} & \approx X^{*}_{\Sigma\textrm{C}}m,\\
X_{\textrm{CH}_{4}} & \approx X^{*}_{\textrm{CH}_{4}}(P,T),
\end{align}
where $X^{*}_{\textrm{CH}_{4}}(P,T)$ is the mole fraction abundance of CH$_{4}$ as a function of pressure and temperature in a solar metallicity gas, as plotted in Figure \ref{fig: abundances}b.  Unlike CO, which shows a strong ($m^2$) dependence on metallicity inside the CH$_{4}$ field, the methane abundance inside the CO field is independent of metallicity because it is proportional to the total carbon abundance ($\Sigma$C) but inversely proportional to the total oxygen abundance ($\Sigma$O).  Note that these simple abundance approximations begin to break down at higher metallicities ($m>30$) as heavy elements become significantly abundant relative to hydrogen and helium.   For high-metallicity cases, mole fraction values can be calculated from solar elemental abundance ratios \citep{lodders2010} with different values of $m$. A more complete description of the equilibrium chemical behavior of CO, CH$_{4}$, and other carbon-bearing species as a function of $P$, $T$, and $m$ in substellar atmospheres can be found in \citet{lodders2002}.

\subsection{Chemical Kinetics and Timescales}
For CO-CH$_4$ quench kinetics we adopt the reaction scheme of \citet{moses2011} and \citet{visscher2011}, who identify two plausible mechanisms for CO$\rightleftarrows$CH$_{4}$ interconversion in substellar atmospheres.  In general, the rate-limiting step for breaking/forming the C--O bond is 
\begin{equation}\label{rxn:CH3OH}
\textrm{CH}_{3}\textrm{OH}  + \textrm{M} \rightleftarrows \textrm{CH}_{3} + \textrm{OH} + \textrm{M},
\end{equation}
when CH$_{4}$ is the dominant C-bearing gas, and
\begin{equation}\label{rxn:CH2OH}
\textrm{CH}_{2}\textrm{OH} + \textrm{H} \rightleftarrows \textrm{CH}_{3} + \textrm{OH},
\end{equation}
when CO is the dominant C-bearing gas.  However, the contribution of either pathway is significant enough that both should be considered when estimating the CO$\rightleftarrows$CH$_{4}$ interconversion rate in substellar atmospheres.  The chemical timescale for CO$\rightarrow$CH$_{4}$ is thus given by
\begin{equation}\label{eqn: CO timescale}
\tau_{\textrm{chem}}(\textrm{CO})=\frac{[\textrm{CO}]}{k_{\ref{rxn:CH3OH}f}[\textrm{CH}_3\textrm{OH}][\textrm{M}]+k_{\ref{rxn:CH2OH}f}[\textrm{CH}_2\textrm{OH}][\textrm{H}]},
\end{equation}
where $k_{\ref{rxn:CH3OH}f}$ and $k_{\ref{rxn:CH2OH}f}$ are the forward reaction  rate constants for reactions (\ref{rxn:CH3OH}) and (\ref{rxn:CH2OH}), respectively.   The chemical timescale for CH$_{4}$$\rightarrow$CO is given by
\begin{equation}\label{eqn: CH4 timescale}
\tau_{\textrm{chem}}(\textrm{CH}_{4})=\frac{[\textrm{CH}_4]}{k_{\ref{rxn:CH3OH}r}[\textrm{CH}_3][\textrm{OH}][\textrm{M}]+k_{\ref{rxn:CH2OH}r}[\textrm{CH}_3][\textrm{OH}]},
\end{equation}
where $k_{\ref{rxn:CH3OH}r}$ and $k_{\ref{rxn:CH2OH}r}$ are the reverse reaction rate constants for reactions (\ref{rxn:CH3OH}) and (\ref{rxn:CH2OH}), respectively, taken from \citet{jasper2007} and reversed to give $k_{\ref{rxn:CH3OH}f}$ and $k_{\ref{rxn:CH2OH}f}$ using the method described in \citet{visscher2011}.  

Using equilibrium abundances for OH, CO, CH$_{3}$, CH$_{4}$,  CH$_{2}$OH and CH$_{3}$OH, the chemical timescales for CH$_{4}$$\rightarrow$CO and CO$\rightarrow$CH$_{4}$ are shown in Figure \ref{fig: timescales} (as contours on a logarithmic scale) as a function of pressure and temperature in a solar-metallicity gas.  We also calculated chemical timescales at metallicities between $[\textrm{Fe/H}] = -2$ and $[\textrm{Fe/H}] = +2$.  Over this range, the timescale for CO destruction, $\tau_{\textrm{chem}}\textrm{(CO)}$, is independent of metallicity, whereas the timescale for CH$_{4}$ destruction as a function of metallicity is given by
\begin{equation}\label{eqn: tchem met}
\tau_{\textrm{chem}}(\textrm{CH}_{4}) \approx \tau^{*}_{\textrm{chem}}(\textrm{CH}_{4})(P,T)m^{-1},
\end{equation}
where $\tau^{*}_{\textrm{chem}}(\textrm{CH}_{4})(P,T)$ is the chemical timescale for CH$_{4}$$\rightarrow$CO as a function of pressure and temperature in a solar metallicity gas (plotted in Figure \ref{fig: timescales}b) and $m$ is the metallicity factor.  Equation (\ref{eqn: tchem met}) demonstrates that higher metallicities result in shorter timescales (i.e., faster reaction kinetics) for CH$_{4}$$\rightarrow$CO chemistry over the range of pressures and temperatures considered here.

\section{Results and Discussion}
\begin{figure*}
\begin{center}
\scalebox{0.45}{\includegraphics[angle=0]{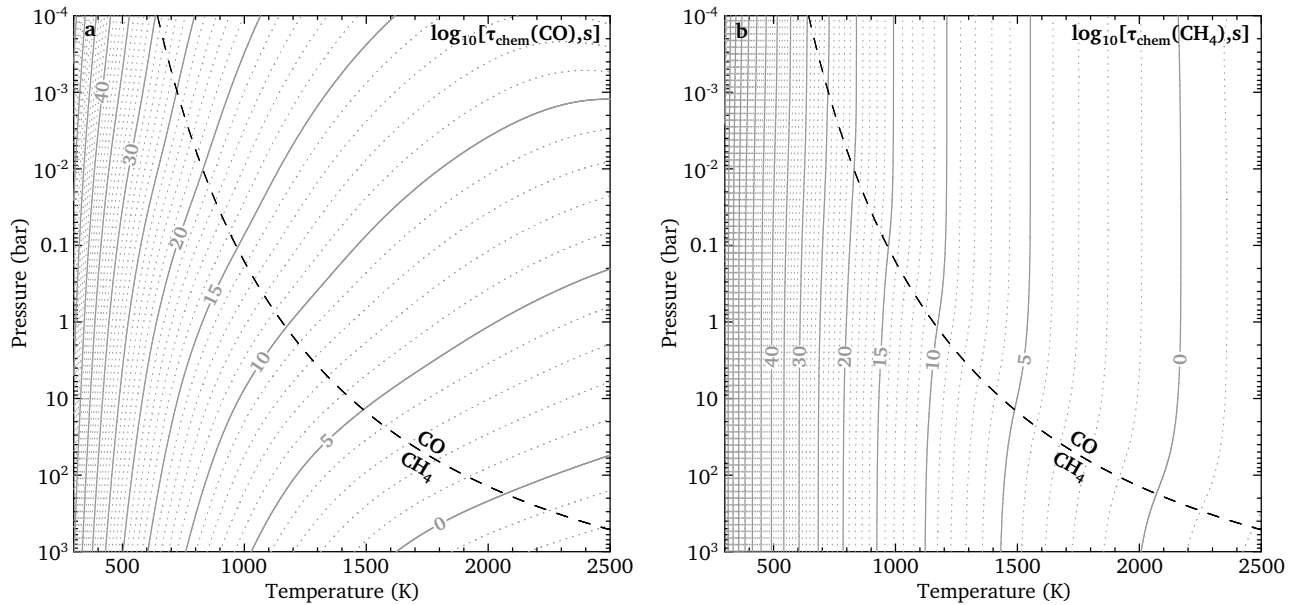}} \end{center} \caption[Blank Timescales]{Chemical timescales (in seconds, on a logarithmic scale) for (a) the conversion of CO to CH$_{4}$ and (b) the conversion of CH$_{4}$ to CO in a solar-metallicity gas.  The dashed line denotes the position of the CO = CH$_{4}$ equal abundance boundary.  Figure after \citet{lodders2002}.  See text for details.}
\label{fig: timescales} 
\end{figure*}

The abundances in Figure \ref{fig: abundances} and the chemical timescales in Figure \ref{fig: timescales} can be compared with orbital and dynamical timescales as well as atmospheric pressure-temperature profiles for individual objects in order to estimate the quench level and abundances of disequilibrium species.  Using CO-dominated HAT-P-2b and CH$_{4}$-dominated CoRoT-10b as specific examples, here we examine various profile/timescale intersections in order to explore CO$\rightleftarrows$CH$_{4}$ quench processes in exoplanet atmospheres, as these intersections represent some quench level with respect to orbital timescales.  The guiding principle throughout the following discussion is that the chemical behavior in the CO$\rightleftarrows$CH$_{4}$ system is governed by the shortest available timescale in the system.

The planet-wide average temperature profiles at periapse and apoapse overlaid on the timescale and abundance plots are shown in Figure \ref{fig: hatp2b} for HAT-P-2b and Figure \ref{fig: corot10b} for CoRoT-10b.  The pressure-temperature profiles for HAT-P-2b are planetary average profiles derived from the general circulation models (GCMs) of \citet{lewis2010,lewis2012}.  The CoRoT-10b profiles are from the one-dimensional models of \citet{fortney2006dyn,fortney2008,fortney2010} where the incident stellar radiation is assumed to be redistributed over the entire planet.  

Because we are primarily interested in the temperature difference between periapse and apoapse (and its effect on thermochemistry), we focus on the time elapsed between these extremes irrespective of radiative timescales.  However, orbit-induced temperature variations will exist only if the radiative timescale is less than the orbital period \citep[e.g., see][and references therein]{cowan2011}.  Orbital timescales for the highly eccentric exoplanets listed in Table \ref{tab: exoplanets} are on the order of days, whereas radiative timescales at photospheric altitudes on such planets are typically on the order of hours \citep[e.g.,][]{laughlin2009,iro2010,lewis2012}. We thus adopt $0.5p$ (where $p$ is the orbital period) as the characteristic orbital timescale throughout the following discussion, and time elapsed between periapse and apoapse is denoted in Figures \ref{fig: hatp2b} and \ref{fig: corot10b} by the lines labeled $0.5p$.  The time elapsed between noon and midnight is denoted by $0.5p_{s}$, which is one half of the pseudo-synchronous rotation period derived from \citet{hut1981} as presented in \citet{iro2010}.  Although we focus on orbital temperature variations in the present study, in principle diurnal variations in atmospheric chemistry could be explored using a similar approach, depending upon the relative dynamical and radiative timescales.

\subsection{Transport-induced Quenching}

In addition to orbit-induced temperature variations, atmospheric constituents will also be subject to convective vertical mixing.  The mixing timescale ($\tau_{\textrm{mix}}$) is estimated using the expression
\begin{equation}
\tau_{\textrm{mix}}=L^{2} / K_{zz},
\end{equation}
where $L$ is the mixing length and $K_{zz}$ is the eddy diffusion coefficient \citep{stone1976,smith1998}.  Because the vertical mixing rate in extrasolar giant planets is unknown, $K_{zz}$ is treated as a free parameter in the models.  For hot giant planet atmospheres, it is reasonable to assume $L\sim0.5H$ \citep{visscher2011,smith1998} where $H$ is the atmospheric scale height ($H=RT/\mu g)$.  If required, better estimates of $L/H$ may be calculated on a case-by-case basis using the method of \citet{smith1998}.  The gravity is calculated from the mass and radius of the planet (see Table \ref{tab: exoplanets}) and we assume a mean molecular mass of $\mu\sim 2.4$ g mole$^{-1}$ for an H$_{2}$-dominated, solar-composition atmosphere.  

\begin{figure*}
\begin{center}
\scalebox{0.45}{\includegraphics[angle=0]{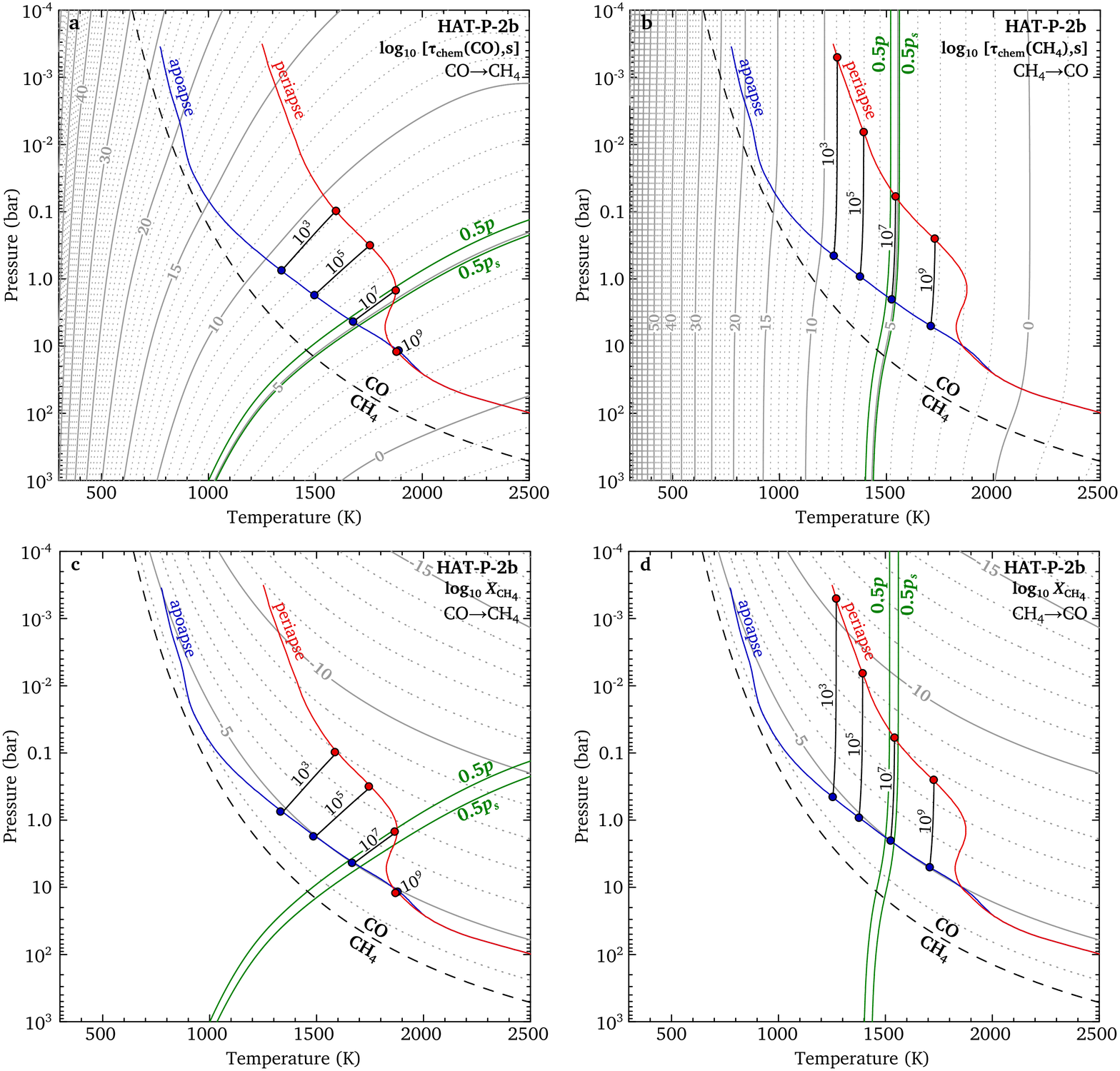}}
\end{center} 
\caption[HAT-P-2b]{Chemical timescales (top panels; contours in seconds on a logarithmic scale) and equilibrium CH$_{4}$ abundances (bottom panels; mole fraction contours on a logarithmic scale) for CO$\rightarrow$CH$_{4}$ conversion timescales (left panels) and CH$_{4}$$\rightarrow$CO conversion timescales (right panels) on HAT-P-2b.  Planet-wide average temperature profiles for HAT-P-2b are shown for apoapse (blue lines) and periapse (red lines).  The solid green lines denote the time elapsed between periapse and apoapse (0.5$p$, where $p$ is the orbital period) and the time elapsed between noon and midnight (0.5$p_s$, where $p_s$ is the pseudo-synchronous rotation rate).  The filled symbols with tie lines show where $\tau_{\textrm{mix}}=\tau_{\textrm{chem}}$ (i.e., the vertical quench level) for different values of $K_{zz}$ (cm$^{2}$ s$^{-1}$) using the mixing length approach of \citet{smith1998} as presented in \citet{visscher2011}.  (A color version of this figure is available in the online journal.)}
\label{fig: hatp2b} 
\end{figure*}

\begin{figure*}
\begin{center}
\scalebox{0.45}{\includegraphics[angle=0]{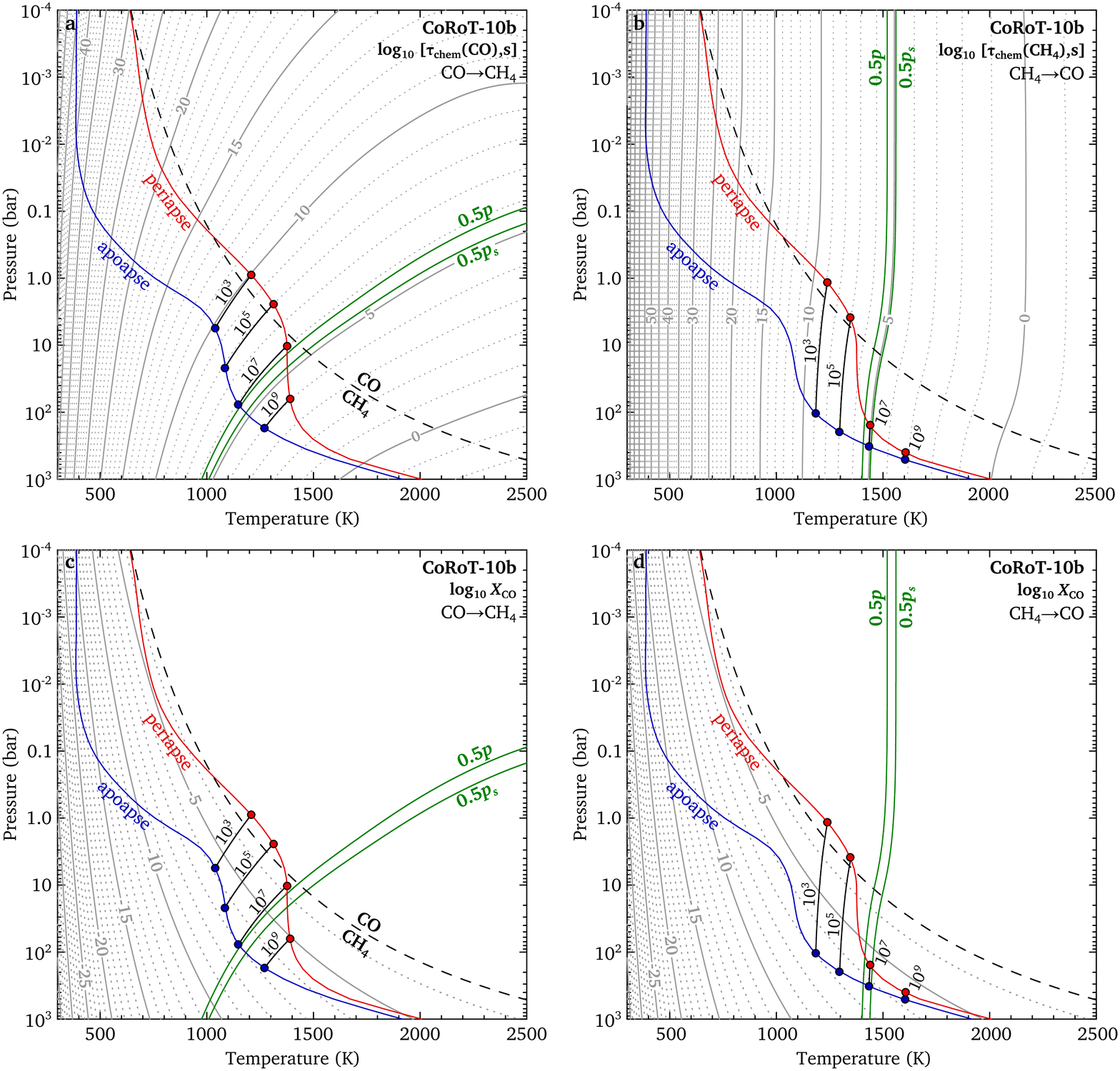}}
\end{center} 
\caption[CoRoT-10b]{Chemical timescales (top panels; contours in seconds on a logarithmic scale) and equilibrium CO abundances (bottom panels; mole fraction contours on a logarithmic scale) for CO$\rightarrow$CH$_{4}$ conversion timescales (left panels) and CH$_{4}$$\rightarrow$CO conversion timescales (right panels) on CoRoT-10b.  Planet-wide average temperature profiles for CoRoT-10b are shown for apoapse (blue lines) and periapse (red lines).  The solid green lines denote the time elapsed between periapse and apoapse (0.5$p$) and the time elapsed between noon and midnight (0.5$p_s$).  The filled symbols with tie lines show where $\tau_{\textrm{mix}}=\tau_{\textrm{chem}}$ (i.e., the vertical quench level) for different values of $K_{zz}$ (cm$^{2}$ s$^{-1}$).  (A color version of this figure is available in the online journal.)}
\label{fig: corot10b} 
\end{figure*}

The vertical quench level is defined as the altitude at which $\tau_{\textrm{mix}}=\tau_{\textrm{chem}}$ \citep{prinn1977,fegley1985apj}.  The pressures and temperatures for which this condition occurs on each planet are shown in Figures \ref{fig: hatp2b} and \ref{fig: corot10b} as filled circles with labeled solid lines denoting the estimated $K_{zz}$ value (in cm$^{2}$ s$^{-1}$) for each case.  The intersections of the atmospheric pressure-temperature profiles with the $\tau_{\textrm{mix}}=\tau_{\textrm{chem}}$ curves in Figures \ref{fig: hatp2b} and \ref{fig: corot10b}  indicate the location of  transport-induced quench level ($P_q$, $T_q$) for different assumptions about the vertical mixing rate (characterized by $K_{zz}$).  The quench level represents the lowest pressure (i.e., highest altitude) along the atmospheric profile for which equilibrium can be achieved in the presence of convective vertical mixing; at all pressures $<P_q$, the abundance of CO (for CO destruction) or CH$_{4}$ (for CH$_{4}$ destruction) will remain fixed at the equilibrium abundance achieved at $P_q$, $T_q$.  As expected, high $K_{zz}$ values result in quenching deep in the atmosphere whereas low $K_{zz}$ values give quenching at lower atmospheric pressures.  However, as discussed below, vertical mixing can only play a significant role in the atmospheres of highly eccentric exoplanets over orbital timescales if mixing timescales ($\tau_{\textrm{mix}}$) are shorter than the time elapsed between periapse and apoapse ($0.5p$).

\citet{cooper2006} demonstrated that disequilibrium abundances of CO and CH$_{4}$ in the upper (observable) atmospheres of hot Jupiters are controlled by vertical quenching, 
and that horizontal transport may homogenize these abundances with longitude.  In the present study we therefore do not examine the effects of horizontal transport, and focus instead on the relative roles of vertical quenching caused by convective mixing and thermal quenching caused by orbit-induced temperature variations.  However, we note that strong horizontal winds may serve to homogenize 
disequilibrium abundances from either quenching mechanism (vertical or orbit-induced) in the upper atmospheres of highly eccentric exoplanets.

\subsection{Methane Quenching on HAT-P-2b}

On HAT-P-2b, CO is the dominant carbon-bearing gas and so we consider CH$_{4}$ quench chemistry.  As the planet cools between periapse and apoapse, the equilibrium CH$_{4}$ abundance is predicted to greatly increase (e.g., see Figure \ref{fig: hatp2b}c).  However, equilibrium can only be maintained if CO$\rightarrow$CH$_{4}$ conversion kinetics are fast enough to produce CH$_{4}$ over orbital timescales ($\tau_{\textrm{chem}}\le 0.5p$).  We therefore examine CO$\rightarrow$CH$_{4}$ chemical timescales (left hand panels in Figure \ref{fig: hatp2b}) in order to determine whether the higher CH$_{4}$ abundances predicted by equilibrium can be produced as the atmosphere shifts from high temperatures at periapse to low temperatures at apoapse.

At relatively high pressures and temperatures, deep in the atmosphere, equilibrium can be maintained throughout the entire orbit if chemical timescales are always shorter than orbital timescales ($\tau_{\textrm{chem}} < 0.5p$).  For highly eccentric exoplanets, the lowest pressure for which this condition holds is given by the intersection of the apoapse profile and the orbital timescale ($0.5p$).  On HAT-P-2b, the apoapse/$0.5p$ intersection occurs near the $4$-bar level (see Figure \ref{fig: hatp2b}a), demonstrating that CO$\rightarrow$CH$_{4}$ equilibrium chemistry -- and the corresponding increase in the equilibrium abundance of CH$_{4}$ -- can be maintained throughout the entire orbit only for $P\ga4$ bars.  In the absence of vertical mixing, the equilibrium CH$_{4}$ mole fraction abundance at the $\sim4$-bar level can track eccentricity-induced temperature variations over orbital timescales by shifting from $\sim1$ ppm at periapse to $\sim10$ ppm near apoapse (see Figure \ref{fig: hatp2b}c).  

At lower pressures, however, orbit-induced temperature variations will quench CO$\rightarrow$CH$_{4}$ conversion once $\tau_{\textrm{chem}}>0.5p$ as the planet moves between periapse and apoapse.  Equilibrium compositions achieved during part of the orbit when the atmosphere is warmer ($\tau_{\textrm{chem}}<0.5p$) will quench once $\tau_{\textrm{chem}}>0.5p$ as the atmosphere cools.  This boundary is represented by the $0.5p$ curve between the apoapse and periapse profiles in Figure \ref{fig: hatp2b}a.  Quenched CH$_{4}$ abundances resulting from orbit-induced thermal quenching can likewise be estimated by the position of the orbital timescale ($0.5p$) curve in Figure \ref{fig: hatp2b}c.  Note that the portion of the orbit over which equilibrium is achieved (i.e., $\tau_{\textrm{chem}}<0.5p$) decreases with altitude in the upper atmosphere of HAT-P-2b.  

The intersection between the periapse profile and the $0.5p$ curve represents the highest altitude (i.e., lowest $P$) in the atmosphere at which equilibrium can be achieved at any point during the orbit, with respect to orbital timescales.  In Figure \ref{fig: hatp2b}a, the intersection between periapse and the $0.5p$ curve occurs near the 1-bar level.  At this intersection, CO$\rightarrow$CH$_{4}$   proceeds over orbital timescales because $\tau_{\textrm{chem}}(\textrm{CO}) = 0.5p$.  As temperatures cool from periapse to apoapse (moving left horizontally from the periapse/$0.5p$ intersection in Figure \ref{fig: hatp2b}a), the rate of CO destruction (and hence CH$_{4}$ production) decreases and quenches when $\tau_{\textrm{chem}}(\textrm{CO}) > 0.5p$.  For HAT-P-2B, this suggests orbit-induced thermal quenching of $\sim0.1$ ppm CH$_{4}$ at the $\sim1$-bar level throughout the orbit, far below the equilibrium abundance of $\sim10$ ppm predicted at apoapse temperatures (see Figure \ref{fig: hatp2b}c).  

However, this thermal quenching scenario neglects the role of vertical transport. The relative importance of vertical versus orbit-induced thermal quenching depends upon the transport timescale relative to the orbital timescale.  As can be seen in Figure \ref{fig: hatp2b}, the orbital timescale for HAT-P-2b is roughly equivalent to a mixing timescales characterized by $K_{zz}\sim10^7$ cm$^{2}$ s$^{-1}$.  For $K_{zz} < 10^7$ cm$^{2}$ s$^{-1}$, orbit-induced thermal quenching from periapse to apoapse will likely determine the CH$_{4}$ abundance because vertical mixing timescales exceed the time elapsed between periapse and apoapse ($\tau_{\textrm{mix}} > 0.5p$). Thermal quenching in this scenario is expected to yield very low ($0.01-1$ ppm) CH$_{4}$ abundances in the upper atmosphere throughout the orbit.

If we assume faster vertical mixing rates, vertical quenching will dominate because $\tau_{\textrm{chem}}<0.5p$.  For $K_{zz}>10^7$ cm$^{2}$ s$^{-1}$, transport-induced quenching will give $\sim10$ ppm CH$_{4}$ throughout the upper atmosphere of HAT-P-2b at apoapse (see Figure \ref{fig: hatp2b}c), because thermally-quenched CH$_{4}$  will be overwhelmed by much larger quantities of CH$_{4}$ transported from higher pressures along the apoapse profile.  As noted by \citet{lewis2012}, even higher CH$_{4}$ abundances may be possible if HAT-P-2b has a reduced metallicity and/or an enhanced C/O ratio \citep[e.g., see][]{moses2012}.  Estimates of the quenched CH$_{4}$ abundance and the expected quench mechanism are summarized in Table \ref{tab: results}.  The large difference in predicted abundances as a function of $K_{zz}$ suggest that observations of the CH$_{4}$ abundance on highly eccentric exoplanets may yield information about atmospheric mixing rates.

\subsection{Carbon Monoxide Quenching on CoRoT-10b}

On the cooler CoRoT-10b, CH$_{4}$ is the dominant C-bearing gas and so we consider CO quench chemistry.  Unlike CH$_{4}$ chemistry on HAT-P-2b, the equilibrium CO abundance on CoRoT-10B is expected to \emph{decrease} as the planet moves from periapse to apoapse (see Figure \ref{fig: corot10b}c), leading to quenching of excess CO.  We thus first consider CO$\rightarrow$CH$_{4}$ kinetic timescales as the planet nears apoapse.  The intersection of the apoapse profile and orbital timescale ($0.5p$) occurs near the $90$-bar level in Figure \ref{fig: corot10b}c, indicating that CO$\rightarrow$CH$_{4}$ equilibrium chemistry can be maintained throughout the orbit if $P\ga90$ bars.  

As the planet swings back toward periapse, the equilibrium abundance of CO is predicted to rapidly increase with increasing temperatures.  However, CO can only be produced if CH$_{4}$$\rightarrow$CO kinetics (Figure \ref{fig: corot10b}b) are sufficiently fast relative to orbital timescales.  We must therefore examine CH$_{4}$$\rightarrow$CO chemical timescales to determine the highest altitude (lowest $P$) at which equilibrium can be maintained throughout the orbit.  The intersection between the apoapse profile and the $0.5p$ curve for CH$_{4}$$\rightarrow$CO timescales occurs near the 300-bar level on CoRoT-10b: i.e., only for atmospheric pressures greater than $\sim300$ bars can CO$\rightleftarrows$CH$_{4}$ equilibrium chemistry be maintained over orbital timescales, shifting from $\sim0.3$ ppm CO at apoapse to $\sim3$ ppm CO at periapse (see Figure \ref{fig: corot10b}d).  However, note that CO abundances in the upper atmosphere of CoRoT-10b at apoapse will more likely be determined by orbit-induced thermal quenching or vertical quenching, depending upon the rate of vertical mixing in its atmosphere.

\begin{deluxetable}{cr@{ }lr@{ }l}
\tablecaption{Estimated Quench Abundances at $P<1$ bar}
\tablewidth{0pt}
\tablehead{
$K_{zz}$ & \multicolumn{2}{c}{HAT-P-2b} & \multicolumn{2}{c}{CoRoT-10b}\\ (cm$^2$ s$^{-1}$) & \multicolumn{2}{c}{CH$_{4}$} & \multicolumn{2}{c}{CO} }
\startdata
10$^3$ & 0.01-0.1&ppm ($t$) & $\ge$100&ppm ($t$)\\
10$^5$ & 0.03-0.1&ppm ($t$) & $\ge$100&ppm ($t$)\\
10$^7$ & 10&ppm ($v$) &  100&ppm ($v,t$)\\
10$^9$ & 10&ppm ($v$) &  10&ppm ($v,t$)\\[-1.5mm]
\enddata
\tablecomments{The notations ($v$) and ($t$) denote whether the quenching mechanism is predominantly caused by vertical mixing or by orbit-induced thermal variations, respectively.}
\label{tab: results}
\end{deluxetable}

The intersection between periapse and the $0.5p$ curve represents the lowest pressure level at which equilibrium can be achieved at any point during the orbit with respect to orbital timescales.  For CoRoT-10b, the intersection between the periapse profile and the $0.5p$ curve  occurs near $14$ bars for CO$\rightarrow$CH$_{4}$ chemistry (Figure \ref{fig: corot10b}a).  At pressure levels between the apoapse/$0.5p$ intersection ($\sim90$ bars) and the periapse/$0.5p$ intersection ($\sim14$ bars), CO$\rightarrow$CH$_{4}$ conversion is quenched whenever $\tau_{\textrm{chem}}>0.5p$ (this boundary is represented by the $0.5p$ curve between the apoapse and periapse profiles in Figure \ref{fig: corot10b}b).  From a comparison of orbital timescales ($0.5p$ curve) and CO abundances in Figure \ref{fig: corot10b}c, we thus might expect thermally-quenched CO abundances of $\sim0.1-100$ ppm above the $\sim90$-bar pressure level on CoRoT-10b.

For example, orbit-induced thermal quenching at the $\sim14$-bar level (the periapse/$0.5p$ intersection) would yield $\sim100$ ppm CO throughout the orbit, in excess of the equilibrium abundance of $\sim1$ ppm CO predicted at apoapse.  This will also overwhelm the $\sim0.1-1$ ppm CO predicted from vertical quenching over a range of $K_{zz}$ values ($10^3-10^7$ cm$^2$ s$^{-1}$).  Once present at lower pressure levels, CO cannot readily be converted back into CH$_{4}$ because $\tau_{\textrm{chem}}(\textrm{CO})>0.5p$ (i.e., it is quenched) and $\tau_{\textrm{chem}}(\textrm{CO})>\tau_{\textrm{chem}}(\textrm{CH}_{4})$ (i.e., the CO production rate exceeds the CO destruction rate).  Although CO production via CH$_{4}$$\rightarrow$CO is also quenched at these altitudes, high abundances of CO are maintained because there is no way to destroy it thermochemically. In addition, as noted above, rapid horizontal transport would tend to homogenize this disequilibrium abundance \citep[e.g.,][]{cooper2006}.  For slower mixing rates ($K_{zz}<10^7$ cm$^2$ s$^{-1}$) it thus appears plausible that thermal quenching from eccentricity-induced temperature variations determines the CO abundance in the upper atmosphere of CoRoT-10b.  

As for HAT-P-2b, the orbital timescale for CoRoT-10b is roughly equal to vertical mixing timescales characterized by $K_{zz}\sim10^7$cm$^2$ s$^{-1}$, and we therefore expect vertical mixing to play a role for $K_{zz}\ga 10^7$ cm$^2$ s$^{-1}$.  Note that for these cases, vertical quenching along the periapse profile may determine a CO abundance which subsequently undergoes thermal quenching at lower pressures.  For example, quenching at $K_{zz}\sim10^7$ or $\sim10^9$ cm$^2$ s$^{-1}$ at periapse would yield quenched CO abundances of $\sim100$ or $\sim10$ ppm, respectively, throughout much of the upper atmosphere.  These quantities of CO would prevail via orbit-induced thermal quenching as the planet moves from periapse to apoapse and overwhelm  disequilibrium CO abundances that result from vertical quenching ($\tau_{\textrm{chem}}=\tau_{\textrm{mix}}$) along the apoapse profile.   

Estimates of the upper atmospheric ($P<1$ bar) quenched CO abundance and the expected quench mechanism(s) are summarized in Table \ref{tab: results}.  Although the results for CoRoT-10b do not show large differences in abundance as a function as $K_{zz}$, they consistently predict disequilibrium CO abundances 2-4 orders of magnitude in excess of equilibrium abundances predicted at apoapse. 

\subsection{Nitrogen Quench Chemistry}   
Other chemical systems may also respond to eccentricity-induced temperature variations.  For example, large changes in atmospheric temperature may shift equilibrium chemistry predictions from relatively higher N$_{2}$ abundances at periapse to relatively higher NH$_{3}$ abundances at apoapse.  To explore this effect, we calculated chemical timescales for N$_{2}$$\rightleftarrows$NH$_{3}$ interconversion using the updated reaction mechanism described in \citet{moses2011}.  For highly eccentric exoplanets, the quench process of most interest is N$_{2}$$\rightarrow$NH$_{3}$ as the planet swings from periapse to apoapose (where higher NH$_{3}$ abundances are predicted).  However, compared to the CO$\rightleftarrows$CH$_{4}$ system,  N$_{2}$$\rightleftarrows$NH$_{3}$ interconversion generally quenches at higher pressures, deeper in the atmosphere, where orbit-induced temperature variations are smaller.  We thus expect orbit-induced thermal quenching to play a relatively minor role on HAT-P-2b and CoRoT-10b, yielding variations in the NH$_{3}$ abundance of less than an order of magnitude.  Vertical quenching will more likely determine the abundance of NH$_{3}$ in the upper atmospheres of these highly eccentric exoplanets.

\section{Summary and Conclusions}

Equilibrium abundances for CO and CH$_{4}$ and chemical timescales for CO$\rightleftarrows$CH$_{4}$ interconversion were calculated as a function of pressure, temperature, and metallicity.  A comparison of the abundance and timescale plots with atmospheric pressure-temperature profiles can be used to estimate the quench levels and abundances of CO and CH$_{4}$ for a variety of substellar objects.  In principle, this approach can be applied for any substellar object with an H$_{2}$-dominated atmosphere that is subject to eccentricity-induced temperature variations.  Overall, orbit-induced thermal quenching tends to favor CO over CH$_{4}$ because CO is the higher-temperature species and because chemical reactions proceed more rapidly at periapse than at apoapse.  Moreover, equilibrium can be maintained to higher altitudes on warmer CO-dominated objects than on cooler CH$_{4}$-dominated objects, with respect to orbital and vertical mixing timescales.

Whether vertical quenching or orbit-induced thermal quenching governs disequilibrium CO$\rightleftarrows$CH$_{4}$ chemistry depends upon the orbital timescale relative to the mixing timescale (as a function of $K_{zz}$) along the atmospheric profile.  In some cases, vertical quenching along the periapse profile may govern the disequilibrium abundances of CH$_{4}$ and CO throughout the upper atmosphere over the entire orbit.  For both HAT-P-2b and CoRoT-10b, the effect of orbit-induced thermal quenching is roughly equivalent to transport-induced quenching assuming $K_{zz}\sim10^{7}$ cm$^{2}$ s$^{-1}$.  For lower $K_{zz}$ values ($<10^{7}$ cm$^{2}$ s$^{-1}$), thermal quenching may have a significant effect, whereas quenching via vertical transport will determine disequilibrium abundances of CO and CH$_{4}$ at higher $K_{zz}$ values ($>10^{7}$ cm$^{2}$ s$^{-1}$).  For CO$\rightleftarrows$CH$_{4}$ chemistry, differences in the quenching mechanism may result in large differences in the abundances of disequilibrium carbon-bearing species.  Refinements to atmospheric structure models and improved observational estimates of CO and CH$_{4}$ on highly eccentric exoplanets may yield important clues to the chemical and dynamical behavior of their atmospheres.    Further development of the chemical models, including the consideration of photochemical production and loss rates throughout the orbit, may likewise provide improved abundance estimates of disequilibrium species throughout the upper atmospheres of highly eccentric transiting exoplanets.

\begin{acknowledgements}
Thanks to J.~Moses and the anonymous referee for helpful comments on the manuscript, to J.~Moses and N.~Lewis for interesting discussions about orbital variations and quench chemistry, and to N.~Lewis and J.~Fortney for providing atmospheric profiles.  This research was supported by the NASA Planetary Atmospheres Program (NNX10AF64G) and has made use of the Exoplanet Orbit Database
and the Exoplanet Data Explorer at exoplanets.org.
\end{acknowledgements}

\end{document}